\documentclass[sigconf]{acmart}
\AtBeginDocument{%
 }
\usepackage{graphicx}  
\usepackage{booktabs}  
\usepackage{colortbl}  
\usepackage{array} 
\usepackage{multirow} 
\usepackage[linesnumbered,ruled,vlined]{algorithm2e} 

\AtBeginDocument{%
  }

\copyrightyear{2024}
\acmYear{2024}
\setcopyright{rightsretained}
\acmConference[AISec '24]{Proceedings of the 2024 Workshop on Artificial Intelligence and Security }{October 14--18, 2024}{Salt Lake City, UT, USA}
\acmBooktitle{Proceedings of the 2024 Workshop on Artificial Intelligence and Security (AISec '24), October 14--18, 2024, Salt Lake City, UT, USA}
\acmDOI{10.1145/3689932.3694756}
\acmISBN{979-8-4007-1228-9/24/10}

\makeatletter
\gdef\@copyrightpermission{
  \begin{minipage}{0.3\columnwidth}
   \href{https://creativecommons.org/licenses/by/4.0/}{\includegraphics[width=0.90\textwidth]{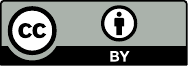}}
  \end{minipage}\hfill
  \begin{minipage}{0.7\columnwidth}
   \href{https://creativecommons.org/licenses/by/4.0/}{This work is licensed under a Creative Commons Attribution International 4.0 License.}
  \end{minipage}
  \vspace{5pt}
}
\makeatother

\begin{document}

\title{Efficient Model Extraction \\via Boundary Sampling}
\subtitle{AISec'24 - Best Paper Award}


\author{Maor Biton Dor}
\orcid{0009-0001-0159-0984}
\affiliation{%
  \institution{Ben-Gurion University}
  \department{Software and System Engineering}
  \city{Be'er Sheva}
  \country{Israel}}
\email{bitondor@post.bgu.ac.il}

\author{Yisroel Mirsky}
\orcid{0000-0001-6367-2734}
\affiliation{%
  \institution{Ben-Gurion University}
  \department{Software and System Engineering}
  \city{Be'er Sheva}
  \country{Israel}}
\email{yisroel@post.bgu.ac.il}








\begin{abstract}
    This paper introduces a novel data-free model extraction attack that significantly advances the current state-of-the-art in terms of efficiency, accuracy, and effectiveness. Traditional black-box methods rely on using the victim's model as an oracle to label a vast number of samples within high-confidence areas. This approach not only requires an extensive number of queries but also results in a less accurate and less transferable model. In contrast, our method innovates by focusing on sampling low-confidence areas (along the decision boundaries) and employing an evolutionary algorithm to optimize the sampling process. These novel contributions allow for a dramatic reduction in the number of queries needed by the attacker by a factor of 10x to 600x while simultaneously improving the accuracy of the stolen model. Moreover, our approach improves boundary alignment, resulting in better transferability of adversarial examples from the stolen model to the victim's model (increasing the attack success rate from 60\% to 82\% on average). Finally, we accomplish all of this with a strict black-box assumption on the victim, with no knowledge of the target's architecture or dataset.

    We demonstrate our attack on three datasets with increasingly larger resolutions and compare our performance to four state-of-the-art model extraction attacks.
\end{abstract}





\maketitle

\section{Introduction}
    Machine learning (ML) models have become an invaluable form of intellectual property. Their development can consume massive resources: extensive datasets, painstaking labeling efforts, substantial computational power, and the expertise of skilled engineers. Consequently, organizations have a critical need to safeguard their ML models from unauthorized replication, as the theft of a well-trained model represents a significant loss of investment.
    
    However, the risks of model theft extend beyond financial considerations. A stolen model exposes its decision boundaries, giving adversaries a blueprint for highly targeted attacks. If attackers steal a model, they can perform a white box attack and craft an adversarial example – a subtly altered input that changes the model prediction \cite{demontis2019adversarial}. For example, consider the case of a facial recognition system used for authentication. With white box access to the model, an attacker can craft an adversarial example of an unauthorized individual which will fool the deployed model and grant access.

    In black-box model extraction, attackers lack internal insights into the target model $f$, including access to its parameters, architecture, or training data ($\mathcal{D}_0$). To extract the model’s functionality, attackers use $f$ as an oracle to label a new dataset $\mathcal{D}_1$, capturing the decisions of $f$. They then train a substitute model $f'$ on $\mathcal{D}_1$ aiming to replicate the decision boundaries of $f$ \cite{addepalli2020degan,aivodji2019gamin,barbalau2020black,orekondy2019knockoff,tramer2016stealing,truong2021data,yu2020cloudleak,wang2022dualcf}.

    
    Traditional black-box approaches exhibit two fundamental shortcomings:

    \begin{description}
        \item[Challenges in High-Confidence Sampling:] These methods often prioritize samples where the victim model $f$ demonstrates high classification confidence. The underlying assumption is that these high-confidence samples are representative of the original training distribution in $\mathcal{D}_0$.  However, training exclusively on such data leaves considerable room for error in exactly where the substitute model places its decision boundaries. This harms both the accuracy of $f'$ and its boundary alignment with $f$. Poor boundary alignment means adversarial examples designed to deceive $f'$ are less likely to successfully transfer to the victim model $f$ \cite{demontis2019adversarial}. 
        \item[The High Cost of Query-Based Model Extraction:] Existing techniques are markedly inefficient in terms of the number of queries required, posing a significant issue for black box adversaries who aim to minimize query count to reduce their likelihood of being detected. For instance, extracting even relatively small-scale models, such as those trained on datasets like Fashion-MNIST, can necessitate over 10 million queries to the victim model (evaluated later in the paper). This requirement increases exponentially for larger models, like those trained on ImageNet, where the number of queries could potentially surge to hundreds of millions. This high query volume not only escalates the risk of detection but also underscores the need for more query-efficient model extraction methods.
    \end{description}
    
\begin{figure*}
        \centering  
        \includegraphics[width=0.7\textwidth]{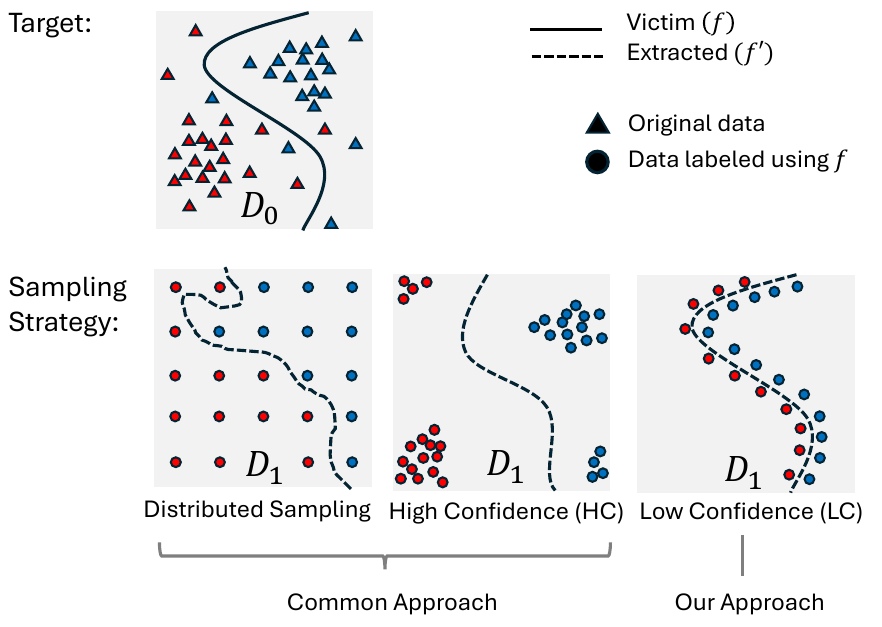}  
        \caption{An illustration that compares sampling strategies used by model extraction attacks. Existing methods that target multi-class classifiers either sample $f$ in a distributed manner or focus on high-confidence areas. Our approach is to focus on low-confidence areas to better capture the decision boundary manifolds.}  
        \label{fig:main}  
    \end{figure*}

    In this paper, we introduce a novel model extraction attack that dramatically improves upon the state-of-the-art in terms of accuracy, efficiency, and effectiveness. We fundamentally shift the focus from high-confidence areas to regions where the victim model exhibits low-confidence (illustrated in Fig. \ref{fig:main}). These low-confidence areas illuminate the model's decision boundaries. By strategically placing labeled samples along these boundaries, we compel the substitute model to closely replicate the victim's behavior. To optimize the efficiency of our sampling process, we employ an evolutionary algorithm. We call this attack \textit{Boundary Aware Model-extraction} (BAM).

    Our approach offers several key advantages:
    \begin{description}
        \item[Reduced Query Count:] By utilizing an evolutionary algorithm, we achieve a dramatic reduction in the number of queries required for model extraction—a factor of 10-600 times less compared to existing methods. This translates to a faster and more efficient attack process for malicious actors. Notably, our approach stands out since the number of queries it needs to steal a deep neural network can be as little as 4.3 times the original dataset's size.
        \item[Improved Model Accuracy:] By focusing on low-confidence regions, our method leads to a stolen model with higher accuracy compared to previous methods. This implies that the stolen model ($f'$) is not merely a poor imitation but a more accurate representation of the original model ($f$). 
        \item[Enhanced Boundary Alignment:] The low-confidence approach also significantly improves the alignment between the stolen model $f'$ and the original model $f$. This translates to a higher success rate when transferring adversarial examples from $f'$ to attack $f$. In simpler terms, the stolen model can be more effectively used to craft targeted attacks against the original model. Our assessments reveal that enhancing alignment increases attack success rate by 37\% on average compared to the current leading approach.
        \item[Data-Free Extraction:] Most prior methods require some knowledge about the target model's training data distribution $\mathcal{D}_0$ to bootstrap their sampling techniques (e.g., to train their generators \cite{barbalau2020black}). Our entire approach operates in a \textbf{strict black-box} setting. This means:
        \begin{enumerate}
            \item The attacker only needs access to the model's query interface to launch the attack.
            \item The attacker has no knowledge of the victim model's data distribution (e.g., that it was trained on images of animals).
        \end{enumerate}
        This significantly lowers the barrier to entry for potential theft.
        \end{description}

    To the best of our knowledge, this is the first model extraction attack that strategically employs an evolutionary algorithm directly on the query generation process, while also targeting low-confidence areas rather than the conventional high-confidence areas. These approaches not only fill a critical scientific gap but also demonstrably achieve state-of-the-art outcomes in both the efficiency and efficacy of model extraction, all while maintaining a \textit{strict} black box setting.
    
    To thoroughly assess our proposed method's effectiveness, we conducted evaluations across various aspects. First, we analyzed the impact of our algorithm's hyperparameters to better understand the method's behavior and to identify the ideal settings. Next, we tested the method on three models with varying levels of complexity, utilizing different datasets for each, and compared its performance against four leading-edge model extraction algorithms. Our proposed methodology outperformed established baselines, achieving an accuracy increase of 4-16\%. This result validates the efficacy of our approach and sets a new benchmark for future research. Furthermore, to demonstrate the robustness of our algorithm, it was tested across various models trained on the same dataset, illustrating its effectiveness across different architectural frameworks. In our final analysis, we assessed the impact of our method on the success of adversarial example transfer attacks, in comparison to outcomes achieved with models extracted via baseline methods. Our approach has notably increased the average attack success rate from an initial 60\% to 82\%, signifying a leap forward from the performance of the closest competing model extraction algorithm. This progress emphasizes the exceptional efficacy of our attack strategy.
    To make our work reproducible, we have published our code online.\footnote{\url{https://github.com/MaorBitonDor/Boundary-Sampling-for-Efficient-Model-Extraction}}

\section{Background}
    This section provides the technical background and notations necessary for discussing this paper.
    
    \textbf{Evolutionary Strategies.} To create training data $\mathcal{D}_1$ for $f'$, we propose the use of an evolution strategy (ES) to efficiently query $f$ for labels. An ES is a method used to solve optimization problems by simulating the process of natural evolution \cite{whitley2001overview}. This algorithm operates on a population of candidate solutions, represented as \(P_t = \{x_1, x_2, \dots, x_n\}\) at iteration \(t\), where each \(x_i\) is an individual solution. The algorithm iteratively applies the principles of selection, crossover, and mutation to generate a new population \(P_{t+1}\) that is hoped to be closer to the optimal solution. Selection chooses individuals based on their fitness, evaluating how well they solve the optimization problem. Crossover combines pairs of selected individuals to produce offspring, introducing new solutions into the population. Mutation randomly alters parts of an individual, introducing variability. Over successive generations, the population evolves towards an optimal or near-optimal solution to the problem. The process repeats until a termination condition is met, such as reaching a maximum number of generations or achieving a solution with satisfactory fitness.

    In this work, we utilize an ES to generate samples over a model's manifolds (decision boundaries). Therefore, we do not consider $x_i \in P_t$ to be a solution, but rather a single sample that must be labeled and added to the dataset $\mathcal{D}_1$. For the same reason, we do not use crossover since the objective is to explore $f$ spatially. There are existing works which study the identification of boundaries \cite{bryan2005active}. In this work, we chose to use an evolutionary strategy since it intuitively explores the entire input space which conducive in making a high-fidelity copy of the model.

    \textbf{Transferability of Adversarial Examples.} An adversarial example can be formally defined as a carefully perturbed input designed to mislead a machine-learning model into making an incorrect prediction. This concept is crucial in understanding the vulnerabilities of models to small, often imperceptible changes in their input data. Let \(x\) be an original input to a model \(f\) that correctly classifies it into class \(y\). An adversarial example \(x'\) is generated by adding a small perturbation \(\delta\) to \(x\), such that \(x' = x + \delta\), with the condition that \(f(x') \neq y\) while ensuring \(\|\delta\|\) is minimal to keep \(x'\) visually indistinguishable from \(x\). The generation of adversarial examples exposes the fragility of deep neural networks in high-dimensional spaces, a phenomenon thoroughly discussed in seminal works such as Szegedy et al. \cite{szegedy2013intriguing} and Goodfellow et al. \cite{goodfellow2014explaining}.

    Transferability, in the context of adversarial examples, refers to the property that allows adversarial examples crafted for one model (\(f'\)) to remain effective against another model (\(f\)). This means that an adversarial example \(x'\), generated to deceive the substitute model \(f'\), will also mislead the victim model \(f\) into making an incorrect prediction. Formally, if \(f'(x') \neq y\) and \(f(x') \neq y\), then the adversarial example \(x'\) is considered to have successfully transferred from \(f'\) to \(f\). Transferability is a critical aspect because it enables attackers to exploit models even without direct access to them, by attacking substitute models that are more accessible. The success of these attacks, especially transfer attacks, relies heavily on how well the gradient and boundaries of $f$ align with $f'$ \cite{demontis2019adversarial,jagielski2020high}.

    \textbf{Quality of Extracted Models.} When stealing $f$, it is important to consider how to measure the quality of the stolen model $f'$. In \cite{jagielski2020high} the authors proposed two qualities: model accuracy and model fidelity. Model accuracy is the task performance of $f'$ (e.g., classification accuracy) compared to that of $f$. If the attacker's objective is to use $f'$ to craft adversarial examples for a transfer attack, then accuracy alone is not a good metric since there are many different boundaries that can obtain the same level of accuracy. One way to measure model fidelity is to measure the transferability (attack success rate) of adversarial examples from $f'$ to $f$. Our proposed attack aims to not only increase the accuracy of the model but also its fidelity.

    \textbf{Strict Black-Box Attack.} To clarify our terminology, we consider a model extraction attack a \textbf{strict black box} attack if the approach does not require any knowledge of the dataset's distribution at all (for example, that it contains images of animals). Therefore, we do not consider works that train their attack models on datasets with a similar distribution to $\mathcal{D}_0$ as strict black-box attacks. This is because such attacks possess knowledge of the victim's training data (including its distribution), which aids in selecting a proxy dataset.

\section{Related Works}
    Historically, knowledge distillation was one of the first techniques to achieve results similar to model extraction. Our work aims to advance the current state-of-the-art in the domain of black-box model extraction. Therefore, we will review relevant topics. A summary of black-box model extraction algorithms for multi-class classifiers can be found in Table \ref{tab:comparison-table}.

    \textbf{Knowledge Distillation.} It has been shown that it is possible to transfer a model's functionality to a simpler model \cite{hinton2015distilling,phuong2019towards}. This process is called knowledge distillation and involves a larger, more complex teacher model and a smaller, more efficient student model which copies the teacher. The concept was originally introduced to mitigate the computational demands of deploying large models on resource-constrained devices. Traditionally the model owner undertakes the distillation process and uses the original dataset used to train the student model. However, some researchers have developed distillation techniques that utilize substitute datasets, which approximate the original dataset's feature space or distribution \cite{barbalau2020black,gou2021knowledge}. Furthermore, there has been a push towards data-free knowledge distillation methods, which eliminate the need for any substitute dataset \cite{truong2021data}.
      
    \textbf{Basic Model Extraction.}
    Black-box settings traditionally employ knowledge distillation strategies, where the victim model serves as an oracle to generate a substitute dataset $\mathcal{D}_1$ for training the extracted model $f'$. The initial approach to model extraction, as described by Tramer et al. (2016) \cite{tramer2016stealing}, formulated equation-solving problems based on the confidence scores provided by the API. These equations relate to the model's parameters, allowing an attacker to solve for these parameters and effectively reconstruct the model. While this method showed promise for compact models, it encountered significant inefficiencies when applied to larger models, primarily due to the vast quantity of queries needed to naively fill the feature space. Additionally, this approach relies on assumptions about the data distribution (e.g., that $\mathcal{D}_0$ consists of images of cats and trucks) and requires knowledge of the teacher's architecture.
    
    \textbf{Generative Approaches.} To mitigate the issue of naive sampling, researchers turned to guided approaches for creating $\mathcal{D}_1$. For example, techniques such as reinforcement learning and active learning have been used to sample the most relevant regions to minimize the number of queries \cite{orekondy2019knockoff,pal2020activethief,jagielski2020high,wang2022black}. However, these approaches assume access to the data in $\mathcal{D}_0$ or at least data that has the same distribution as $\mathcal{D}_0$.
    
    More recently, researchers have moved towards using generative techniques to help augment $\mathcal{D}_1$ to better capture the behaviors of $f$. The common approach is to use a Generative Adversarial Network (GAN) to generate the queries that capture the surface (outputs) of $f$ \cite{addepalli2020degan,aivodji2019gamin,barbalau2020black,kariyappa2021maze,truong2021data}. For instance, the DeGAN framework effectively achieves this by initially utilizing a proxy dataset akin to \(\mathcal{D}_0\). It then employs this dataset to train a GAN with the goal of generating images to form $\mathcal{D}_1$. These images are designed to mimic the distribution of \(\mathcal{D}_0\) through the process of querying the function \(f\). 
    To improve the quality of the extracted models further, Barbalau et al. \cite{barbalau2020black} filtered the data given to $f'$ by using an evolutionary search. Their approach was to first train a GAN $g$ to generate an image $x$ based on a model's confidence on that image $h(x)$, where $h$ is a classifier trained by the attacker. Next, they create $\mathcal{D}_1$ by sending millions of random samples to the victim in the form of $g(z)$, where $z$ is random noise. Finally, $f'$ is trained on the highest confidence samples for one epoch and the process repeats by sending another batch to $f$. 

    While the use of GANs to generate $\mathcal{D}_1$ can yield high-quality models, the training process is highly overt and very inefficient requiring 10s to 100s of millions of queries on $f$ to either train the GAN or collect a substantial dataset. In contrast, our approach achieves even higher quality models with only a fraction of the number of queries (0.6-1.4 million). Moreover, the evolutionary strategy we propose significantly differs from that of \cite{barbalau2020black} in two ways: (1) we optimize the query images directly as opposed to going through the generator in a GAN and (2) we use the algorithm to explicitly find the boundaries of $f$ whereas they use it to find a random cover of the high-confidence regions of $f$. The result is that our approach is significantly more efficient and effective at copying the parameters of $f$, and it is easier to implement as well. 
      
    \textbf{Data Free Approaches.} 
    To make model extraction work in stricter black box settings, researchers have looked into ways to perform the attack without assumptions on the distribution of $\mathcal{D}_0$. In works such as \cite{addepalli2020degan,barbalau2020black,kariyappa2021maze} $f$ is not queried using an auxiliary dataset, but rather random images created by a generator. 
    Another example is CloudLeak \cite{yu2020cloudleak} where the model extraction algorithm uses a pre-trained model to sample the decision boundaries of $f$ using active learning.

    However, these works are not strict black box attacks because the auxiliary models used in these methods are trained on datasets that are specifically selected to match $\mathcal{D}_0$. The work of Truong et al. \cite{truong2021data} does provide a strict black-box approach but trains a GAN during the attack raising the query count significantly. 
    
    Lastly, of all the methods we reviewed, we found that they either focus on creating a $\mathcal{D}_1$ that fits the victim's entire loss surface or just high-confidence areas alone. In contrast, we argue and demonstrate that fitting to low-confidence areas yields (1) a higher quality model in terms of performance and (2) a higher fidelity model in terms of attack transferability.

    \begin{table*}[t]  
        \centering  
        \caption{Comparison of Different Model Extraction Methods. HC and LC mean high-confidence and low-confidence respectively. $\mathcal{D}_0$ is the victim's training data. Entries with a "-" are cases where the method was evaluated with access to $\mathcal{D}_0$ but could \textit{possibly} be done without.}  
        \label{tab:comparison-table}  
        \resizebox{0.7\textwidth}{!}{%
            \begin{tabular}{@{}lcccccc@{}}  
                \toprule  
                \textbf{Algorithm} & \multicolumn{3}{c}{\textbf{Query Algorithm}} & \multicolumn{2}{c}{\textbf{Black Box Assumptions}} \\  
                \cmidrule(lr){2-4} \cmidrule(lr){5-6}  
                & \textbf{GAN} & \textbf{Sampling} & \textbf{Query} & \textbf{Access} & \textbf{Knowledge} \\  
                & \textbf{Based} & \textbf{Strategy}& \textbf{Rate} & \textbf{to \(\mathcal{D}_0\)} & \textbf{of \(\mathcal{D}_0\)} \\  
                \midrule  
                \cite{tramer2016stealing} API & No & Both & High & No & Yes \\  
                \cite{orekondy2019knockoff} Knockoff Nets & No & Both & Medium & No & Yes \\  
                \cite{pal2020activethief} ActiveThief & No & Both & Medium & No & Yes \\  
                \cite{aivodji2019gamin} Gamin & Yes & HC & Medium & No & Yes \\  
                \cite{yu2020cloudleak} CloudLeak & No & LC & Low & - & Yes \\  
                \cite{jagielski2020high} HA\&HF & No & HC & Low & - & Yes \\  
                \cite{kariyappa2021maze} Maze & Yes & HC & High & No & Yes \\  
                \cite{wang2022dualcf} DualCF & No & HC & Medium & No & Yes \\  
                \cite{barbalau2020black} Black Box Ripper (BBR) & Yes & HC & Very High & No & Yes \\  
                \cite{addepalli2020degan} DeGAN & Yes & HC & Medium & No & Yes \\  
                \cite{truong2021data} DFME & Yes & HC & Medium & No & No \\  
                \rowcolor{gray!25} BAM (ours) & \textbf{No} & \textbf{LC} & \textbf{Low} & \textbf{No} & \textbf{No} \\  
                \bottomrule  
            \end{tabular}}  
    \end{table*}
    \setlength\tabcolsep{6pt} 

\section{Method}

\subsection{Objective}\label{subsec:objective}

Our primary objective is to construct a dataset \(\mathcal{D}_1\) comprising labeled samples that closely follow the decision boundary manifolds of a black box model \(f\), thereby enabling the training of a substitute model \(f'\) that replicates the behavior of $f$ with high fidelity, which is crucial for successful transfer attacks. More formally the problem can be stated as follows: 

\vspace{0.5em}
\noindent\textbf{Problem Statement.} \textit{Given a black-box target classifier \(f\) with no assumptions on its input range or the distribution of its training data $\mathcal{D}_0$, efficiently generate a dataset \(\mathcal{D}_1 = \{(x_1, y_1), (x_2, y_2), \ldots \)\}, where each sample \(x_i\) is positioned near a manifold of \(f\), and \(y_i\) is the label assigned by \(f(x_i)\).  
}\vspace{0.5em}

The decision boundary manifolds represent regions within the input space where the model \(f\) exhibits low-confidence in its predictions, indicating the transition zones between classes. 
For example, consider a multi-class classification scenario where an input sample \(x\) is predicted by the model \(f\) to belong to one of \(C\) classes. The decision boundary manifold between any two classes \(c_i\) and \(c_j\) can be conceptually located where \(f\) assigns prediction confidence approximately equal for these two classes and significantly lower for all other classes. Mathematically, this can be expressed as follows: 

Let \(f(x) = \{p_1, p_2, \ldots, p_C\}\) to be the predicted probability distribution over \(C\) classes for a sample \(x\), where each \(p_i\) represents the model's confidence that \(x\) belongs to class \(i\). A sample \(x\) is said to lie on the decision boundary manifold between classes \(c_i\) and \(c_j\) if $p_i \approx p_j \approx \frac{1}{2}$ and $p_k \approx 0 \quad \forall k \notin \{i, j\}$. Generalizing, a sample \(x\) lies on a manifold involving \(n\) classes if $\sum_{i=1}^{n} p_i \approx 1$ with $p_i \approx \frac{1}{n}$ for the involved classes and $p_k \approx 0$ for all other classes.
The challenge lies in efficiently identifying and sampling these low-confidence manifolds to construct \(\mathcal{D}_1\). 

\subsection{Approach}
\label{sec:approach}
Our solution involves identifying regions within the input space of $f$ that exhibit low-confidence by using an ES: We start with $N$ random samples as generation $P_0$. These samples are selected with a wide range to ensure that they span the entire feature space. Next, we repeat the following for $I$ iterations: 
\begin{enumerate}
    \item \textbf{Selection.} We pass $P_t$ through $f$ to measure the fitness of the samples in $P_t$ using their probabilities (where higher fitness indicates lower confidence), and to collect their labels. We then select the top $k$ samples that have the highest fitness (i.e., samples closest to a boundary) as $S_t$.
    \item \textbf{Mutation.} We mutate the selected samples in $S_t$ by adding scaled random noise to them, creating a new population $P_{t+1}$ of $N$ samples. The previous generation $P_t$ is then added to $D_1$.
\end{enumerate}

This ES transitions $P_0$ from a broadly distributed set of samples across the feature space to an increasingly focused set of samples which lay on the manifolds of $f$. When the samples approximate these manifolds, the introduced mutations (or noise) facilitate exploration along them. Consequently, \(D_1\) accumulates samples that meticulously outline the manifold surfaces, capturing the decision boundaries of $f$.

We will now discuss our implementation of the selection and mutation steps in more detail. The complete pseudo-code for the proposed model extraction attack can be found in Algorithm \ref{alg:attack}.

\subsubsection{Selection} 
Our fitness function is designed to evaluate how close a sample is to a manifold based on the output of the black-box model \( f \). It is defined as 
\begin{equation}\label{eq:fitness}
    \text{Fitness}(x) = -\max(f(x)) 
\end{equation}
where \( f(x) \) represents the vector of predicted probabilities for all classes for the sample \( x \). The rationale behind this definition is that \( \max(f(x)) \) captures the highest confidence $f$ has. Following the intuition in Section \ref{subsec:objective}, we prefer samples that have a lower $\max(f(x))$ since they will be closer to a decision manifold. Therefore, we take the negative of this value so that fitter samples in $P_t$ will be the ones with the lowest confidence. In other words, $\text{Fitness}(x_i)>\text{Fitness}(x_j)$ if $\max(f(x_i))<\max(f(x_j))$ for $\forall x_i,x_j \in P_t$.


\begin{algorithm}[t]
\SetAlgoLined
\DontPrintSemicolon
\KwIn{Black-box model $f$, number of generations $I$, population size $N$, selection size $k$, mutation scale $\gamma$}
\KwOut{Model $f'$ that copies the performance and fidelity of $f$}
Initialize $P_0$ with $N$ random samples (noise)\;
Initialize $\mathcal{D}_1 \gets \emptyset$\;
\For{$t = 0$ to $I-1$}{
    $Y_t \gets f(P_t)$\; 
    $S_t \gets \text{Sort } P_t \text{ by } \text{Fitness}(x) \text{ in descending order, using } Y_t$\;
    $S_t \gets S_t[0:k]$\; 
    Initialize $P_{t+1} \gets \emptyset$\;
    Calculate span $w$ for the features in $P_t$\;
    \ForEach{$x \in S_t$}{
        \For{$i = 1$ to $\frac{N}{k}$}{
            Add $\text{Mutate}_{\gamma}(x)$ to $P_{t+1}$\;
        }
    }
    Add $P_t$ and its labels $Y_t$ to $\mathcal{D}_1$\;
}
Train substitute model $f'$ on dataset $\mathcal{D}_1$\;
\Return{$f'$}\;
\caption{BAM: Boundary Aware Model Extraction}\label{alg:attack}
\end{algorithm}

\subsubsection{Mutation} 
To mutate a sample $x$, we add random uniform noise to its features. However, since we are operating in a strict black box setting, we do not know how much noise is considered reasonable for each feature in $x$. Therefore, we scale the amount of noise according to the min-max values of the samples in $P_t$. Doing so defers this decision to the ES algorithm during the selection step. 

The span of the \(i\)-th feature across all samples in \(P_t\) is calculated as \(w_i = \max((P_t)_{:,i}) - \min((P_t)_{:,i})\), where \((P_t)_{:,i}\) denotes the vector of values for the \(i\)-th feature across all samples in \(P_t\). With this, we define the mutation of a sample as:
\begin{equation}
        \text{Mutate}_{\gamma}(x) = x + \gamma z
\end{equation}
where \(z\) is a vector with each element \(z_i \sim U(-w_i, w_i)\) drawn from a uniform distribution specific to the span \(w_i\) of the \(i\)-th feature, and \(\gamma\) is a parameter of the algorithm that scales the amount of mutation. Based on our experiments, we found \(\gamma = \frac{1}{10}\) to provide the best results.
To create a population of $N$ samples for $P_t$, we mutate each of the selected samples a number of times.

\subsubsection{Boundary Traversal}
When running BAM, we observed that the new generations create samples that seek out manifolds between more classes. This results in a traversal of the model's boundaries which helps us in our attack. The reason this occurs is because the fitness function only penalizes the class with the highest confidence which causes exploration along manifold \textit{towards} manifolds between more classes. 

To explain the intuition, consider a sample with a high confidence on one class $c_i$: if this sample is selected for mutation then some of its children will have a slightly lower confidence on $c_i$ and they will be selected in the next round. If we proceed in this manner, then eventually we will reach a point where the children will have equal confidence in two classes $c_i$ and $c_j$. These samples lie on the manifold between these classes. The children of these generations will either have a larger $c_i$ or $c_j$ depending on the mutation. The mutations affect all classes, so the max operation in (\ref{eq:fitness}) will cause an alternation between the top classes and cause the subsequent generations to follow the $c_i/c_j$ manifold until a third class reaches the same confidence, and so on. 

Since the first population of BAM is generated at random, most of the samples will be classified by $f$ with high confidence on some class $c\in C$. As a result, future generations will seek manifolds of 2 classes, then 3 and so on until the class confidence reaches an equilibrium or a local minimum is reached due to limitations in the mutation step.

    \section{Experiment Setup}

        \subsection{Datasets} 
        For our experiments, we utilized three diverse datasets with increasing resolutions: Fashion-MNIST (1x28x28) \cite{xiao2017fashion}, CIFAR-10 (3x32x32) \cite{krizhevsky2009learning}, and a dataset of 10 Monkey Species (3x224x224) \cite{montoya2018monkey}. For all datasets, we adhered to their published standard train-test splits, utilizing the test portion to evaluate the performance of both the victim ($f$) and the stolen model ($f'$).

        \subsection{Model Extraction Algorithms} 
        We compared our method to four state-of-the-art black box model extraction attacks which also operate on soft labels: \textit{Data-Enriching GAN} (DeGAN) \cite{addepalli2020degan}, \textit{Black-Box Ripper} (BBR) \cite{barbalau2020black}, \textit{CloudLeak} (CL) \cite{yu2020cloudleak} and \textit{Data-Free Model Extraction} (DFME) \cite{truong2021data}.
        
        These attacks employ a variety of techniques: DeGAN creates a dataset to indirectly instruct $f'$ on mimicking the behavior of $f$, whereas DFME employs a GAN to directly compel $f'$ to replicate the actions of $f$. BBR operated similarly to both but uses an ES to select the best samples for training $f'$ (as opposed to using the ES to guide the search like our method). CL employs active learning strategies to generate adversarial samples positioned near the victim model's decision boundaries. Finally, DeGAN, BBR and CL have some light assumptions on the distribution of $\mathcal{D}_0$ but DFME is a strict black-box attack like ours. For DeGAN, BBR, and DFME attacks, we used the author's published code and used there default hyperparameters. For CloudLeak (CL), we had to implement the algorithm ourselves since no working public code is available. We carefully replicated their approach based on the descriptions provided in their paper. We implemented two versions of the CL algorithm for comparison:

        \begin{itemize}
            \item \textbf{CL-Aux}: This variant follows the evaluation setup in the original CL paper; employing an auxiliary dataset for generating samples for sampling. This version is not a strict black-box attack like BAM.
            \item \textbf{CL-Random}: In our pursuit to establish CL as a strict black box attack, we have introduced the CL-Random variant. This approach leverages a dataset comprised of random samples for the sampling process. These samples are generated using the torch.rand(image\_shape) function. Each element of the tensor is a random number drawn from a uniform distribution in the range [0, 1).
        \end{itemize}

        In both the CL-Aux and CL-Random variants, the dataset size was set to 1 million images for the Fashion-MNIST and CIFAR-10 datasets, and 2 million images for the Monkey dataset.

        For DeGAN, we use two different learning rates as done by the authors in their evaluation (lr=0.001 and lr=0.0001). Although DeGAN was originally proposed as a model distillation method, the authors also employed it for model extraction, and their publicly available code uses it in this context. We utilized DeGAN similarly, training the student model using the teacher model as a black-box.

        For our method (BAM) we used the following hyperparameters unless otherwise noted: For CIFAR-10 and Fashion-MNIST, we used a population size $N$ of 20,000 and a selection size $k$ of 6,000. The algorithm was run for 30 and 40 iterations respectively (until convergence). For Monkey, we set $N=\text{10,000}$ and $k=\text{3,000}$ for 140 iterations.

        \subsection{Models}
        For each dataset, we used a different set of architectures for $f$ and $f'$.    
        To demonstrate a black box setting, we ensured that the architectures of $f$ and $f'$ were different. The exception is for the Monkey dataset since we followed the experiment setup of BBR \cite{barbalau2020black}. In table \ref{tab:cifar10-auc-acc-model-table}, we evaluate more combinations of models and show that our method maintains superiority regardless of the victim's architecture.

        For Fashion-MNIST, we used an 11-layer CNN classifier for $f$ and an uninitialized ResNet-18 architecture for $f'$. For CIFAR-10, $f$ was the pre-trained AlexNet from \cite{barbalau2020black}, while $f'$ was again an uninitialized ResNet-18 architecture. For the Monkey dataset, both $f$ and $f'$ were based on the TorchVision ResNet-18 architecture, with $f$ trained on the Monkey dataset and $f'$ left uninitialized.


        \subsection{Resources} 
        In our experiments, we utilized a system equipped with an Intel i7-11700K CPU and 128GB of RAM. For the GPU, we deployed NVIDIA GeForce RTX 3090 graphics cards, leveraging their substantial computational power to train all our models. The time taken to query $f$ to create $\mathcal{D}_1$ varied significantly across datasets. It took 2 min, 3 min and 60 min to run the query portion of the attack on Fashion-MNIST, CIFAR-10 and Monkey respectively.
    
        To train models on $\mathcal{D}_1$,  it took 2 hours, 4 hours, and 2 days for Fashion-MNIST, CIFAR-10 and Monkey respectively.

        \subsection{Metrics}
            In our evaluation, we adopted the metrics of Accuracy, Area Under the Curve (AUC), and Attack Success Rate (ASR) based on the methodology outlined in \cite{jagielski2020high}. These metrics are critical for assessing the effectiveness of model extraction methods and for comparing the performance of our approach against established benchmarks.
            
            \begin{itemize}
                \item \textbf{Accuracy \& AUC}: We use accuracy to measure how well the stolen model performs compared to the victim model. We also consider AUC; a metric that measures the overall predictive capability of a model where an AUC of 1.0 indicates a perfect classifier and an AUC of 0.5 indicates a model that is guessing at random.
                
                \item \textbf{ASR}: To measure model fidelity, we consider the ASR of a basic transfer adversarial attack; adversarial examples are generates using the stolen model and then evaluated on the victim model. Here, a higher ASR suggests that the loss surface and boundaries of the stolen model are more aligned with the victim model.
            \end{itemize}

        \subsection{Experiments}       
        In this paper, our objectives were twofold: firstly, to extract the functionality of a given model $f$, more \textit{efficiently} and \textit{effectively} than current methodologies allow; and secondly, to demonstrate the viability of using the extracted model $f'$ for adversarial example transfer attacks. These aims underpin our experimental approach, designed to both validate the efficiency of our model extraction process and its utility in adversarial contexts.
        
        To this end, we conducted a series of four experiments:
        \begin{enumerate}
            \item \textbf{Task Performance:} We assessed our model extraction attack's performance against the established baselines, utilizing the metrics of accuracy (ACC) and the area under the receiver operating characteristic curve (AUC). Here, the performance of each surrogate model $f'$ was benchmarked against the performance of the original model $f$, serving as the upper limit of achievable extraction fidelity.
        
            \item \textbf{Transfer Attacks:} We evaluated how well the extracted substitute models ($f'$) facilitate transfer attacks. For this purpose, we used the test sets to create adversarial examples. The examples were generated on $f'$ and evaluated on $f$. Performance was measured in terms of attack success rate (ASR), which gauges the proportion of adversarial examples that successfully deceive the victim model. To generate the adversarial examples, we used PGD \cite{madry2017towards} with an $\epsilon=\frac{30}{255}$ over an $\ell_\infty$ norm for 10 iterations. 
            
            \item \textbf{Ablation Study:} To substantiate our claim that targeting low-confidence (LC) regions (boundary manifolds) is more effective than focusing on high-confidence (HC) areas or a combination of both (LC \& HC), we performed an ablation study. In this study, we modified our algorithm to focus on HC regions (by making (\ref{eq:fitness}) positive) and evaluated the model's performance when trained on HC data and when trained on both HC and LC data. 
            
            \item \textbf{Hyperparameters:} To understand the influence of our hyperparameters, we investigated how selection size $k$, population size $N$, and the number of evolutionary iterations $I$ affect performance.
        \end{enumerate}

         \begin{table*}[t]
            \centering
            \caption{Image Modality: The classification performance of models extracted using different attack algorithms. The top row is the performance of the victim model $f$ (the upper bound).}
            \label{tab:auc-acc-table}
            \begin{tabular}{@{}lcccccc@{}} 
            \toprule
            \textbf{Algorithm} & \multicolumn{2}{c}{\textbf{Fashion-MNIST}} & \multicolumn{2}{c}{\textbf{CIFAR-10}} & \multicolumn{2}{c}
            {\textbf{Monkey}} \\ \cmidrule(l){2-7} 
             & ACC & AUC & ACC & AUC & ACC & AUC \\ \midrule
            Victim $f$ & $0.931$ & 0.9935 & $0.825$ & 0.9813 & $0.9$ & 0.9884 \\ \hline
            DFME & $0.585 \pm 0.01$ & 0.92 & $0.58 \pm 0.01$ & 0.92 & $0.617 \pm 0.02$ & 0.92 \\
            DeGAN lr=1 & $0.62 \pm 0.01$ & 0.94 & $0.69 \pm 0.01$ & 0.94 & $0.14 \pm 0.02$ & 0.57 \\
            DeGAN lr=2 & $0.71 \pm 0.01$ & 0.957 & $0.7 \pm 0.01$ & 0.945 & $0.15 \pm 0.02$ & 0.61 \\
            BBR & $0.83 \pm 0.01$ & 0.96 & $0.71 \pm 0.01$ & 0.956 & $0.63 \pm 0.02$ & 0.932 \\
            DisGuide & $0.22 \pm 0.01$ & 0.65 & $0.36 \pm 0.01$ & 0.73 & $0.19 \pm 0.02$ & 0.63 \\
            CL-Random & $0.7485 \pm 0.01$ & 0.958 & $0.19 \pm 0.01$ & 0.597 & $0.12 \pm 0.02$ & 0.562 \\
            CL-Aux & $0.671 \pm 0.01$ & 0.947 & $0.3622 \pm 0.01$ & 0.745 & $0.12 \pm 0.02$ & 0.575 \\
            \rowcolor{gray!25} BAM (ours) & $\textbf{0.8627} \pm 0.01$ & \textbf{0.985} & $\textbf{0.7464} \pm 0.01$ & \textbf{0.964} & $\textbf{0.735} \pm 0.01$ & \textbf{0.961} \\ 
            \bottomrule
            \end{tabular}

                \centering
                \vspace{2em}

                \caption{Image Modality: Comparative analysis across more cases where the victim uses a different architecture for $f$. In all cases, the adversary uses a modified ResNet18 for $f'$.}
                \label{tab:cifar10-auc-acc-model-table}
                \begin{tabular}{@{}lcccccccc@{}} 
                \toprule
                \textbf{Algorithm} & \multicolumn{2}{c}{\textbf{Lenet}} & \multicolumn{2}{c}{\textbf{ResNet9}} & \multicolumn{2}{c}{\textbf{DenseNet}} & \multicolumn{2}{c}{\textbf{Query}} \\ 
                \cmidrule(lr){2-3} \cmidrule(lr){4-5} \cmidrule(lr){6-7} \cmidrule(lr){8-9}
                 & ACC & AUC & ACC & AUC & ACC & AUC & \multicolumn{2}{c}{Count} \\ 
                \midrule
                Victim $f$ (upper bound) & $0.83$ & 0.976 & 0.808 & 0.9795 & 0.7706 & 0.969 & \multicolumn{2}{c}{-} \\
                DFME & $0.35 \pm 0.01$ & 0.702 & $0.36 \pm 0.01$ & 0.746 & $0.56 \pm 0.01$ & 0.904 & \multicolumn{2}{c}{10M} \\
                DeGAN lr=1 & $0.51 \pm 0.01$ & 0.91 & $0.52 \pm 0.01$ & 0.903 & $0.57 \pm 0.01$ & 0.91 & \multicolumn{2}{c}{12.21M} \\
                DeGAN lr=2 & $0.52 \pm 0.01$ & 0.915 & $0.52 \pm 0.01$ & 0.901 & $0.58 \pm 0.01$ & 0.918 & \multicolumn{2}{c}{12.21M} \\
                BBR & $0.55 \pm 0.01$ & 0.921 & $0.48 \pm 0.01$ & 0.886 & $0.53 \pm 0.01$ & 0.902 & \multicolumn{2}{c}{200M} \\
                DisGuide & $0.32 \pm 0.01$ & 0.683 & $0.34 \pm 0.01$ & 0.706 & $0.41 \pm 0.01$ & 0.804 & \multicolumn{2}{c}{20M} \\
                \rowcolor{gray!25} BAM (ours) & $\textbf{0.63} \pm 0.01$ & \textbf{0.932} & $\textbf{0.55} \pm 0.01$ & \textbf{0.914} & $\textbf{0.61} \pm 0.01$ & \textbf{0.922} & \multicolumn{2}{c}{\textbf{1M}} \\
                \bottomrule
                \end{tabular}
        \vspace{1em}
        \end{table*}

    \section{Experiment Results}
    \subsection{Task Performance}
    The classification performance of the stolen image models is presented in Table \ref{tab:auc-acc-table}. The results demonstrate that our method significantly outperforms the current state-of-the-art. Compared to the second-best attack (BBR), our method achieves a 4-5\% higher accuracy on the lower-resolution datasets (Fashion-MNIST and CIFAR-10) and a 16.6\% higher accuracy for the higher-resolution dataset (Monkey). Although BBR is only a few percent away from our method, BBR requires between 86 and 600 million queries to achieve these results, whereas our method only necessitates between 0.6 and 1.4 million queries (see Table \ref{tab:query-count-table}). When considering more efficient attacks, such as DFME and DeGAN, these methods require approximately 10 to 12 million queries (about 10 times more than ours) but exhibit significantly lower performance: On the low-resolution datasets, DFME and DeGAN achieve accuracies of around 0.58 and 0.71, respectively, while our method reaches between 0.74 and 0.86. For the higher-resolution dataset, the performance gap is even more pronounced, with DFME achieving 0.61 accuracy compared to 0.73 for our method, while DeGAN fails to extract the model entirely. Similar performance can be seen regardless of which architecture is used by the victim, as shown in Table \ref{tab:cifar10-auc-acc-model-table}.

    The CL attack shares similarities with BAM in that it samples decision boundaries. However, this attack utilizes active learning which relies on pre-trained models to craft examples close to the boundaries. In contrast, our method is a strict black box attack; it does not depend on pre-trained models or make any assumptions about the training data. Both variants of the attack (CL-Aux and CL-Random) performed worse than BAM by a significant margin. The strict black box variant (CL-Random) comes the closest to BAM on Fashion-MNIST with an accuracy of 0.67 compared to BAM's 0.86. However, this performance drops to dramatically to near random guessing on more complex datasets in terms of color (CIFAR-10) and resolution (Monkey). In contrast, BAM maintain performance comparable to the victim model in all cases.


    A critical benchmark in model extraction is the ratio of queries made compared to the size of the original dataset (i.e., $\frac{|\mathcal{D}_1|}{|\mathcal{D}_0|}$). Table \ref{tab:query-count-table} shows that our method can achieve a query ratio of 4.28:1 for the high-resolution dataset, which is very close to the ideal ratio of 1:1. For the lower-resolution datasets, our method has a query ratio of 12.4:1 to 16.4:1, which is still considerably lower than the ratios of other attack methods which range from 200:1 to 12,080:1.

    In summary, our method not only extracts models with superior classification performance but also marks a significant advancement in query efficiency. Unlike the baselines which have a trade-off between query count and performance, our method maintains a lead in \textbf{both metrics} making it a practical and effective strict black box model extraction technique.

    \setlength\tabcolsep{4pt} 
    \begin{table}[t]  
        \centering  
        \caption{The number of queries performed on $f$ by each attack, and the ratio of that volume compared to the original dataset size ($|\mathcal{D}_0|$).}  
        \label{tab:query-count-table}  
        \resizebox{\columnwidth}{!}{%
            \begin{tabular}{@{}lcccccc@{}}  
                \toprule  
                \textbf{Algorithm} & \multicolumn{2}{c}{\textbf{Fashion-MNIST}} & \multicolumn{2}{c}{\textbf{CIFAR-10}} & \multicolumn{2}{c}{\textbf{Monkey}} \\   
                \cmidrule(l){2-7}   
                & \textbf{Queries} & \textbf{Ratio} & \textbf{Queries} & \textbf{Ratio} & \textbf{Queries} & \textbf{Ratio} \\   
                \midrule  
                DFME & 10M & 200.0 & 10M & 200.0 & 9.16M & 27.83 \\  
                DeGAN & 12.21M & 244.2 & 12.21M & 244.2 & 12.21M & 37.10 \\  
                BBR & 410M & 8200.0 & 604M & 12080.0 & 86M & 261.32 \\  
                \rowcolor{gray!25} BAM (ours) & \textbf{0.62M} & \textbf{12.4} & \textbf{0.82M} & \textbf{16.4} & \textbf{1.41M} & \textbf{4.28} \\   
                \bottomrule  
            \end{tabular}  
        }  
    \end{table}  
    \setlength\tabcolsep{6pt} 
       
    \subsection{Transfer Attacks}
     In Table \ref{tab:fidelity-table}, we present the attack success rates (ASR) of adversarial examples on $f$ when generated using $f'$. We found that our method outperforms the transferability of the other extracted models. The average ASR of our extracted model is significantly higher (0.82) compared to those extracted by DFME (0.27), DeGAN (0.43) and BBR (0.60). When examining specific datasets, our method achieves particularly high results on Fashion-MNIST (0.88) and CIFAR-10 (0.90) demonstrating that model extraction can be an effective way to perform black-box adversarial example attacks.
     
    The effectiveness of our method in achieving a high ASR stems from our targeted sampling strategy. By focusing our sampling around the decision boundaries of the model \(f\), a model trained on dataset \(\mathcal{D}_1\) is better equipped to ensure its adversarial examples will also be effective against \(f\). Essentially, if an example \(x'\) is adversarial for the surrogate model \(f'\), it is likely to be so for \(f\) as well. This principle is illustrated in Fig. \ref{fig:main}, where \(f'\) and \(f\) exhibit similar decision boundaries. Nevertheless, identical decision boundaries alone do not ensure a successful attack. This is because many white-box attacks rely on gradient information to locate these boundaries and models trained on \(\mathcal{D}_1\) may not have accurate gradient information outside the boundary regions. Therefore, to enhance transferability further, it would be necessary to sample regions of $f$ that are not low-confidence.

        \setlength\tabcolsep{4pt} 
        \begin{table}[t]  
            \centering  
            \caption{The performance of adversarial examples generated on \(f'\) and executed on \(f\), measured in ASR.}  
            \label{tab:fidelity-table}  
            \resizebox{\columnwidth}{!}{%
                \begin{tabular}{@{}lcccc@{}}  
                    \toprule  
                    & \multicolumn{3}{c}{\textbf{Datasets}} & \multicolumn{1}{c}{\textbf{Average}} \\   
                    \cmidrule(lr){2-4}  
                    \textbf{Algorithm} & \textbf{Fashion-MNIST} & \textbf{CIFAR-10} & \textbf{Monkey} & \textbf{ASR} \\   
                    \midrule  
                    DFME & 0.36 & 0.35 & 0.107 & 0.272 \\   
                    DeGAN lr=1 & 0.35 & 0.85 & 0.092 & 0.431 \\   
                    DeGAN lr=2 & 0.34 & 0.86 & 0.077 & 0.426 \\   
                    BBR & 0.44 & 0.866 & 0.497 & 0.601 \\   
                    \rowcolor{gray!25} BAM (ours) & 0.876 & 0.904 & 0.687 & 0.822 \\   
                    \bottomrule  
                \end{tabular}  
            }  
        \end{table}  
        \setlength\tabcolsep{6pt} 

        \begin{table*}[t]
            \centering
            \caption{Ablation study on our method, analyzing the impact of where we sample $f$: low-confidence regions representing boundaries (LC), high-confidence regions (HC), and a mix of both (HC \& LC).}
            \label{tab:ablation}
            \begin{tabular}{@{}lccccccccc@{}}  
                \toprule
                \multirow{2}{*}{\textbf{Sample Type}} & \multicolumn{3}{c}{\textbf{Fashion-MNIST}} & \multicolumn{3}{c}{\textbf{CIFAR-10}} & \multicolumn{3}{c}{\textbf{Monkey}} \\ 
                \cmidrule(l){2-4} \cmidrule(l){5-7} \cmidrule(l){8-10} 
                & ACC & Query Count & ASR & ACC & Query Count & ASR & ACC & Query Count & ASR \\ 
                \midrule
                LC only & $0.8627$ & 0.62M & 0.876 & $0.7464$ & 0.82M & 0.904 & $0.735$ & 1.41M & 0.687 \\   
                HC only & $0.7388$ & 0.62M & 0.746 & $0.7365$ & 0.82M & 0.889 & $0.683$ & 1.41M & 0.482 \\   
                HC \& LC & $0.8364$ & 1.24M & 0.915 & $0.7648$ & 1.64M & 0.9218 & $0.676$ & 2.82M & 0.584 \\  
                Half HC \&  Half LC & $0.832$ & 0.64M & 0.907 & $0.7438$ & 0.84M & 0.881 & $0.6066$ & 1.42M & 0.436 \\  
                \bottomrule
            \end{tabular}
        \end{table*}

        \setlength\tabcolsep{6pt} 
        \begin{figure*}[t]  
            \centering  
            \includegraphics[width=0.8\textwidth]{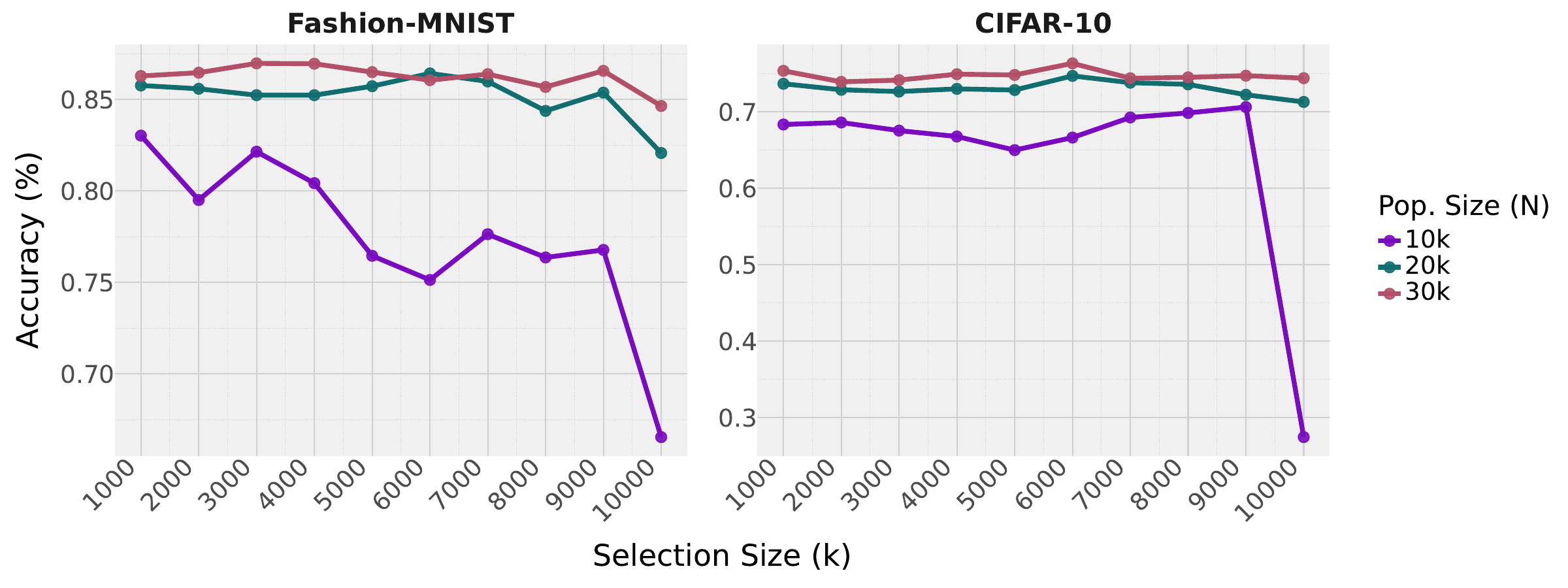}  
            \caption{Comparison of substitute model accuracy across different values of the top k values for Fashion-MNIST and CIFAR-10.}  
            \label{fig:top_k_comparison}  
            \vspace{1em}  
            \centering  
            \includegraphics[width=0.8\textwidth]{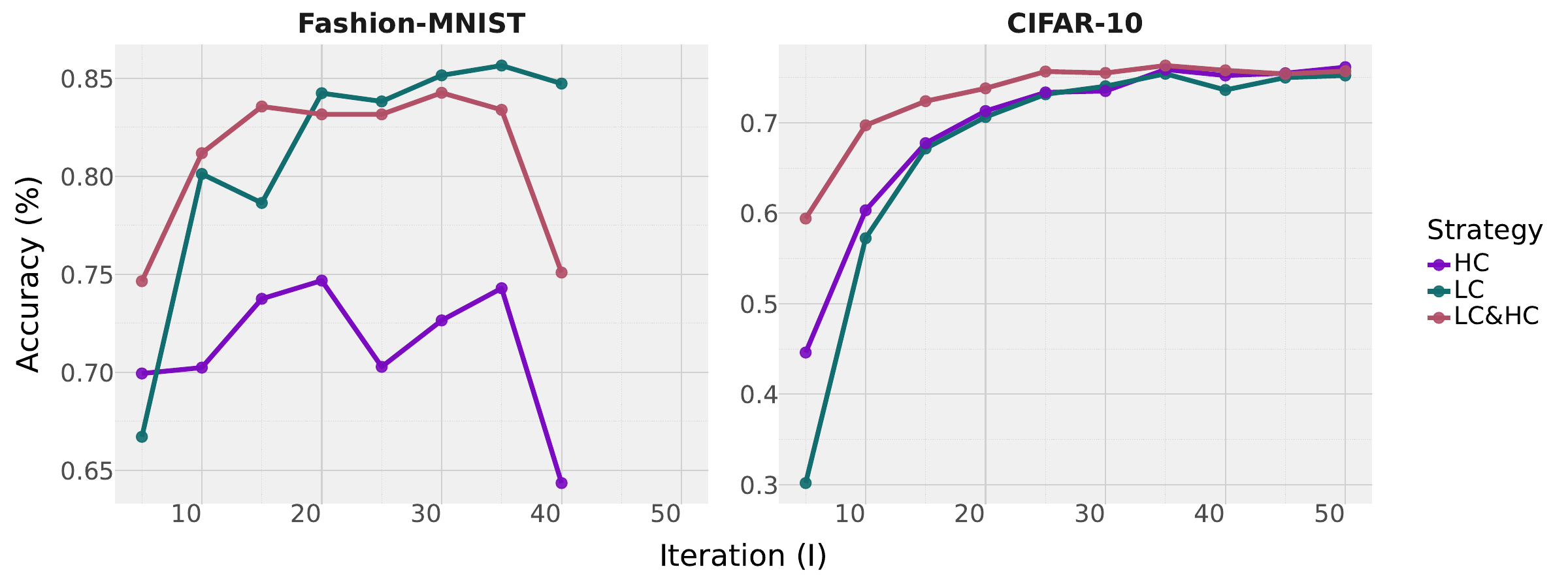}  
            \caption{Comparison of substitute model accuracy across different values of the number of generations for Fashion-MNIST and CIFAR-10.}  
            \label{fig:generation_across_strategy}  
        \end{figure*}

    \subsection{Ablation Study}
    Since our method includes two novel techniques (ES and boundary sampling) it is not clear what is contributing to the improved ACC, ASR and query counts. Therefore, we sampled $f$ with our ES algorithm using two different fitness functions: LC for sampling boundaries of $f$ (defined in (\ref{eq:fitness})) and HC for sampling the data `expected' by $f$ (positive form (\ref{eq:fitness})). 

    Table \ref{tab:ablation} shows that, when we use HC data for $f'$, our performance drops in terms of accuracy. However, it drops to the same level of performance as the next best model extraction technique (BBR in Table \ref{tab:auc-acc-table}). This is because by sampling HC only, we are essentially following the same strategy as existing HC sampling methods. Since the number of queries remains minimal, it is clear that the proposed ES strategy is what's contributing to our algorithm's efficiency and that our boundary sampling strategy (LC) is what's contributing to our enhanced performance. 
    
    Interestingly, when we combine the samples extracted from LC and HC, the ASR improves in most cases. This supports our discussion earlier that white box attacks such as PGD rely on gradient information in other regions to find the boundaries. This also supports the findings of Demontis et al. \cite{demontis2019adversarial} who proposed that the transferability of white box adversarial examples is dependent on how well the gradients (i.e., loss surface) of the two models are aligned.

    In summary, our method is efficient because of the ES and is effective because we sample LC regions. To improve adversarial example transferability, it is possible to run our attack twice, once on for LC and once for HC, to obtain a higher fidelity substitute model, all while benefiting from significantly fewer queries than other state-of-the-art attacks.

    \subsection{Hyperparameters}
    In Fig. \ref{fig:top_k_comparison}, we present the impact of population size $N$ and selection size $k$ on the accuracy of $f'$. Here the number of iterations were fixed to 30 and 40 for Fashion-MNIST and CIFAR-10 respectively. The plots show that $N$ is more important than $k$ so as long as $k<N$ to ensure there is evolution over time. We found that it is possible to use BAM to extract good models with even fewer samples, however the performance will be moderately less.

    In Fig. \ref{fig:generation_across_strategy}, we plot the impact of the number of iterations $I$, using the same $N$ and $k$ from the performance evaluation. The plots reveal that more complex datasets, like CIFAR-10, gain from additional iterations up to a point. This plateau occurs because the algorithm struggles to identify new manifolds post-convergence, a situation that could benefit from algorithmic resets to fresh starting points upon each convergence. Additionally, the HC sampling strategy outperforms LC for the first few iterations. This is because HC can provide simpler expression of $f$ with fewer samples compared to mapping the manifolds of $f$. However, beyond a critical sample volume, the benefits of a well-sampled manifold become more pronounced. Finally, we found that the combination of the HC and LC sampling strategies yields better results, but this is primarily because it effectively doubles the dataset size.

    In summary, LC sampling is the most effective approach for extracting a model. While it is possible to extract $f$ with fewer queries, there is a trade-off with model performance that must be considered.

\section{Discussion \& Limitations}
    In this work, we utilized evolutionary strategies to specifically target LC areas to capture the model's decision boundaries. This approach has proven effective in improving both accuracy and attack success rates in transfer attacks. However, there are some insights and limitations that warrant further discussion.
    
    \textbf{Detection.} Typically, LC samples are rare in typical use cases. Therefore, it may be possible to detect and block BAM if large numbers of LC queries are identified.    
    One way to mitigate this risk is by distributing the queries across multiple users or devices. However, implementing and evaluating the effectiveness of this distributed approach within the context of evolutionary strategies presents a complex challenge that requires further exploration outside of this work. Moreover, BAM is a significant threat in settings where policing model outputs cannot be done safely. For example, when reverse engineering the model of a autonomous vehicle; the car cannot block queries or their could be dangerous consequences in the case of a false positive. 
    
    \textbf{Active Learning.} Interestingly, our results show that LC-based methods can outperform those focusing HC areas. This suggests that LC datasets may be more advantageous for training models. This observation aligns with findings in active learning research, where LC regions are critical for capturing the most relevant information. However, there are scenarios where this attack strategy might underperform—specifically when dealing with overly complex models trained on too few samples. In such cases, the model might introduce boundaries in HC areas according to the victim model, leading to reduced effectiveness of the attack. Therefore, careful selection of population size and model complexity is crucial when applying the BAM approach.

\section{Conclusion}
    In this work, we have introduced a novel approach to model extraction that significantly surpasses the state-of-the-art in efficiency, accuracy, and ability to facilitate transfer-based adversarial attacks. Through our innovative use of evolutionary strategies focused on sampling low-confidence areas near decision boundaries, we have demonstrated a dramatic reduction in the number of queries required to extract a model, while simultaneously improving the fidelity of the extracted model. Our approach,  marks a significant advancement in the field of model extraction by operating under strict black-box assumptions, requiring no prior knowledge of the victim model's architecture or dataset; not even knowledge of the dataset's range of features. 
    
    In conclusion, our findings underscore the potential risks associated with model extraction attacks and highlight the need for developing robust defense mechanisms. As we continue to advance the boundaries of what is possible in machine learning, it is imperative that we also consider the implications of these technologies and strive to protect against their misuse.

\section*{Acknowledgments}
    This work has received support from the Israel National Cyber Directorate, the Zuckerman STEM Leadership Program and funding from the European Union's Horizon 2020 research and innovation program under grant agreement 952172.

\bibliographystyle{ACM-Reference-Format}
\bibliography{sample-base}

\appendix

\end{document}